\documentclass{aa}  

\usepackage{graphicx}
\usepackage{txfonts}
\usepackage{rotating}
\usepackage{color}
\usepackage{soul}
\usepackage{xcolor}
\usepackage{float}
\usepackage{longtable}
\usepackage{lscape}
\usepackage{orcidlink}

\usepackage{amsmath} 
\usepackage{verbatim}
\usepackage[utf8]{inputenc}
\newcommand{\kms}{km~s$^{-1}$}
\newcommand{\msun}{M$_\odot$}
\newcommand{\nic}{$^{56}$Ni}

\begin{document}

   \title{Red eminence: The intermediate-luminosity red transient AT~2022fnm}

   \author{S. Moran \orcidlink{0000-0001-5221-0243}
          \inst{1}
          \and
          R. Kotak \orcidlink{0000-0001-5455-3653}
          \inst{1}
          \and 
          M. Fraser \orcidlink{0000-0003-2191-1674}
          \inst{2}
          \and
          A. Pastorello
          \inst{3}
          \and
          Y.-Z. Cai \orcidlink{0000-0002-7714-493X}
          \inst{4, 5, 6}
          \and
          G. Valerin 
          \inst{7}
          \and 
          S. Mattila 
          \inst{1,8}
          \and
          E. Cappellaro 
          \inst{3}  
          \and T. Kravtsov \orcidlink{0000-0003-0955-9102} 
          \inst{9,1}
          \and C. P. Guti\'errez
          \inst{10,11}
          \and N. Elias-Rosa
          \inst{3, 11}
          \and A. Reguitti \orcidlink{0000-0003-4254-2724}
          \inst{12, 3}
          \and P. Lundqvist \orcidlink{0000-0002-3664-8082}
          \inst{13}
          \and
          A. V. Filippenko \orcidlink{0000-0003-3460-0103} 
          \inst{14}
          \and
          T. G. Brink \orcidlink{0000-0001-5955-2502} 
          \inst{14}
          \and
          X.-F. Wang
          \inst{15, 16}
}

   \institute{Department of Physics and Astronomy, University of Turku, Vesilinnantie 5, FI-20500, Finland\\
              \email{semoke@utu.fi}  
              \and
          School of Physics, University College Dublin, Belfield, Dublin 4, Ireland
        \and INAF -- Osservatorio Astronomico di Padova, Vicolo dell'Osservatorio 5, I-35122 Padova, Italy
          \and Yunnan Observatories, Chinese Academy of Sciences, Kunming 650216, P.R. China
          \and Key Laboratory for the Structure and Evolution of Celestial Objects, Chinese Academy of Sciences, Kunming 650216, P.R. China
          \and International Centre of Supernovae, Yunnan Key Laboratory, Kunming 650216, P.R. China      
          \and Dipartimento di Fisica e Astronomia "G. Galilei", Università degli studi di Padova Vicolo dell'Osservatorio 3, I-35122 Padova, Italy
          \and School of Sciences, European University Cyprus, Diogenes street, Engomi, 1516 Nicosia, Cyprus 
          \and European Southern Observatory, Alonso de Córdova 3107, Casilla 19, Santiago, Chile 
          \and Institut d’Estudis Espacials de Catalunya (IEEC), Gran Capit\`a, 2-4, Edifici
Nexus, Desp. 201, E-08034 Barcelona, Spain
          \and Institute of Space Sciences (ICE, CSIC), Campus UAB, Carrer de Can
Magrans, s/n, E-08193 Barcelona, Spain
          \and INAF - Osservatorio Astronomico di Brera, Via E. Bianchi 46, I-23807, Merate (LC), Italy
          \and Oskar Klein Centre, Department of Astronomy, Stockholm University, Albanova University Centre, SE-106 91 Stockholm, Sweden 
          \and Department of Astronomy, University of California, Berkeley, CA 94720-3411, USA
          \and Physics Department, Tsinghua University, Beijing, 100084, China
          \and Beijing Planetarium, Beijing Academy of Science and Technology, Beijing, 100044, China
    }

   \date{Received ?; accepted ?}

    \abstract{
      We present results from a five-month-long observing campaign of the 
      unusual transient AT~2022fnm, which displays properties common to both luminous red novae (LRNe) and intermediate-luminosity red transients (ILRTs).
      Although its photometric evolution is broadly consistent with that of LRNe, no second peak is apparent in its light curve, and its spectral properties are more reminiscent of ILRTs. It has a fairly rapid rise time of 5.3$\pm$1.5\,d, reaching a peak absolute magnitude of $-12.7\pm$0.1 (in the ATLAS $o$ band).
      We find some evidence for circumstellar interaction, and a near-infrared excess
      becomes apparent at approximately +100\,d after discovery. We attribute this to a dust echo. Finally, from an analytical diffusion toy model, we attempted to reproduce the pseudo-bolometric light curve and find that a mass of $\sim$4~\msun~is needed. Overall, the characteristics of AT 2022fnm are consistent with a weak stellar eruption or an explosion reminiscent of low-energy type IIP supernovae, which is compatible with expectations for ILRTs.
    }

\maketitle
%

\section{Introduction}
With the advent of wide-field surveys over the last two decades, new classes or subgroups of transients have been emerging, necessitating a revision of our understanding of the final stages of stellar evolution. One such grouping occupies a range of luminosities between those typical of supernovae (SNe) and classical novae, and are usually referred to as intermediate-luminosity optical transients (\citealt{Berger2009}) or gap transients \citep[e.g.][]{Kasliwal2012}. Here, we focus on these and the distinction between luminous red novae (LRNe) and intermediate-luminosity red transients (ILRTs).

\begin{figure}[ht]
\includegraphics[width=1\linewidth]{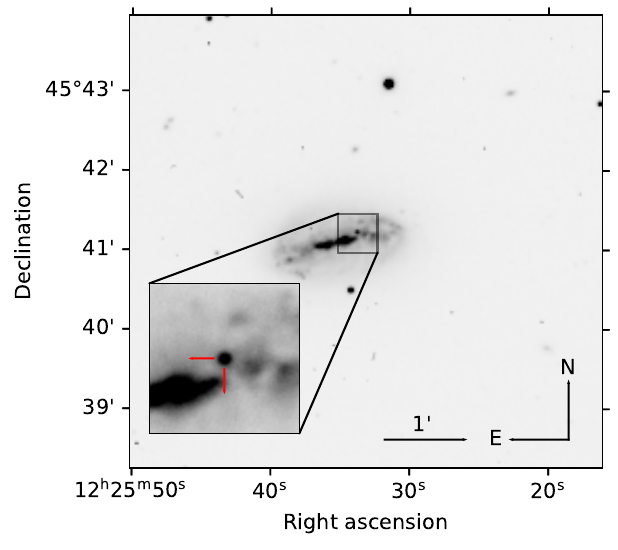}
  \caption{$B$-band image taken at the NOT (ALFOSC) at +3\,d showing the location of AT~2022fnm ($\alpha=12^{\mathrm{h}}25^{\mathrm{m}}33^{\mathrm{s}}.713$, $\delta =+45^{\circ}41'12''.45$, J2000) in its host galaxy, NGC 4389.}
  \label{fig:finder}
\end{figure}
Luminous red novae  have a number of distinctive features. Ranging in luminosity from $-$4 to $-$15~mag, they display double-peaked light curves \citep[e.g.][]{Blagorodnova2017, Pastorello2019a, Pastorello2019b}. The initial peak is thought to result from the cooling emission of the common envelope after its ejection in a stellar-merger scenario \citep{Metzger2017, Blagorodnova2021, Matsumoto2022}. Around the time of the first peak, the spectra are rather blue and display strong Balmer and Fe~{\sc II} emission lines \citep{Cai2019, Cai2022b}. Later, there is a second peak that is often attributed to matter ejection or hydrogen recombination \citep{Pastorello2019a,Matsumoto2022}. By the time of the second peak, the Balmer lines are weaker and a forest of metal lines (mostly in absorption) appears \citep{Pastorello2019b}. At late times, the spectra resemble those of an M-type star, displaying molecular absorption bands such as TiO and VO \citep{Martini1999, Lynch2004, Pastorello2019a, Pastorello2021a}. The strongest evidence for the stellar-merger origin of LRNe comes from V1309~Sco, which showed a decrease in period consistent with an inspiralling binary \citep{Tylenda2011}. 

On the other hand, ILRTs usually rise to peak over about two weeks (though this is based on a limited sample of about ten objects), and their peak absolute magnitudes tend to fall between $-$11.5 and $-$14.5 \citep[e.g.][]{Cai2018,Cai2021,Stritzinger2020b}. They display a post-peak plateau, somewhat reminiscent of type IIP SNe. ILRTs show calcium emission features at all stages of their evolution: in particular, the [Ca~{\sc ii}] $\lambda\lambda$7291,7323 doublet, characteristic of ILRTs, and the $\lambda\lambda\lambda$8498,8542,8662 near-infrared (NIR) Ca~{\sc ii} triplet. There are several proposed mechanisms for the production of ILRTs, including electron-capture SNe from super-asymptotic giant branch star progenitors \citep{Thompson2009, Pumo2009}, massive stars in dusty cocoons experiencing outbursts \citep{Bond2009, Humphreys2011}, and accretion onto a compact object \citep{Kashi2010}.

\begin{figure*}
\sidecaption
\includegraphics[width=12cm]{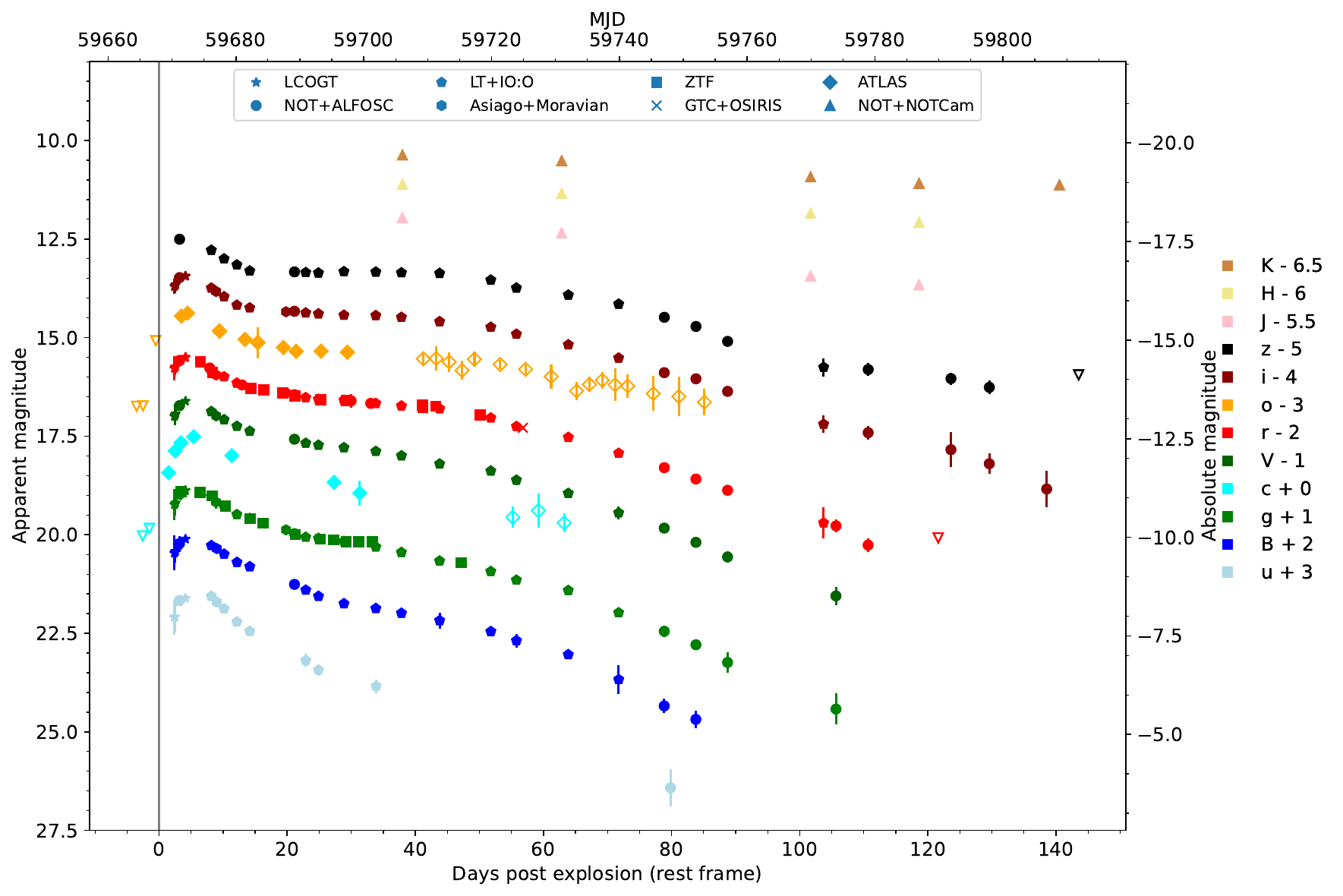}
  \caption{Multi-band light curve of AT~2022fnm. The inverted open triangles represent upper limits. The vertical line marks the epoch of the onset of the outburst. ATLAS data beyond +40\,d are also shown with open symbols as  they are noisier.}
  \label{fig:lc}
\end{figure*}

\begin{figure*}
\sidecaption
\includegraphics[width=12cm]{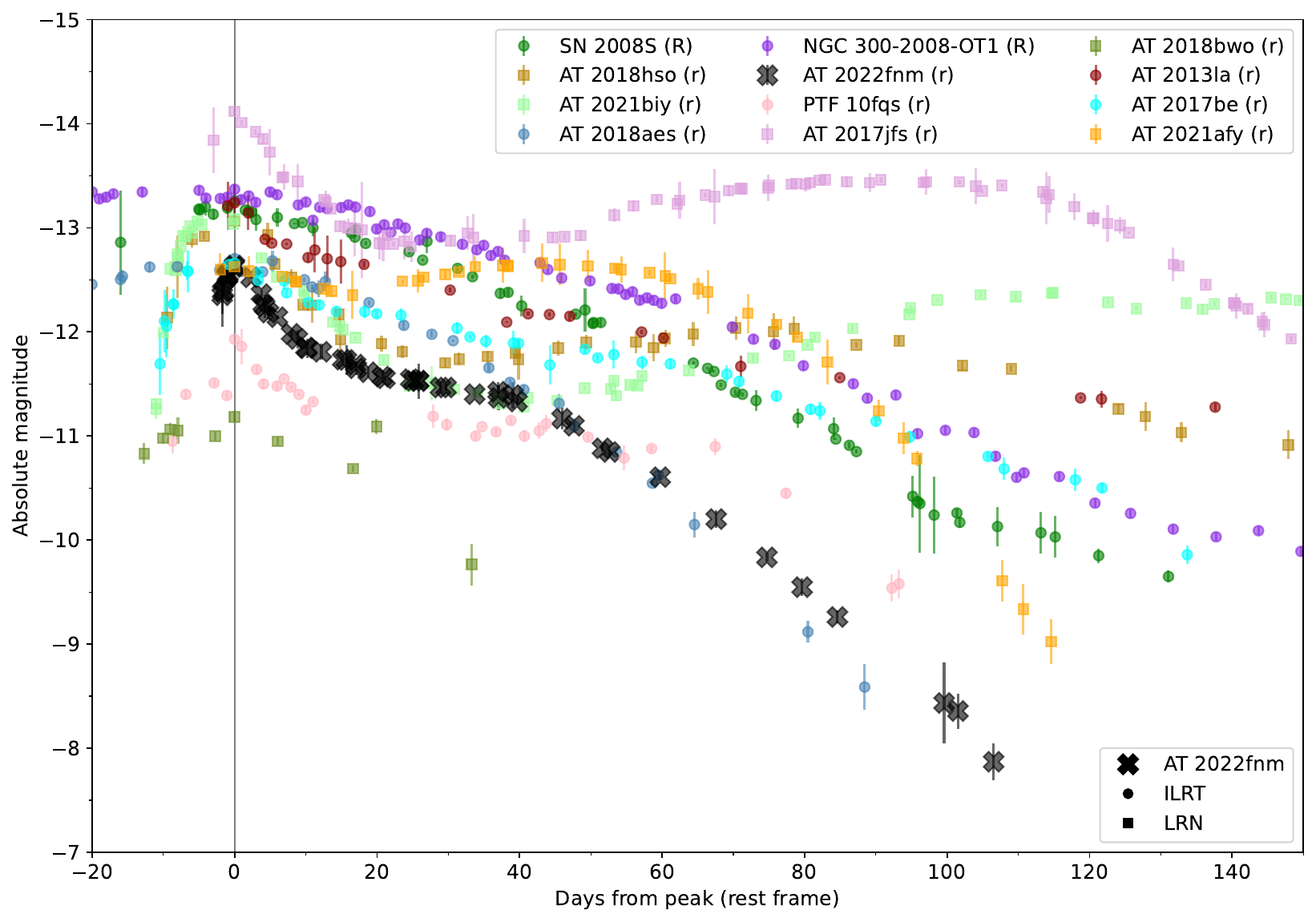}
  \caption{$r$-band comparison of AT~2022fnm with other ILRTs and LRNe from the literature. All objects have been corrected for reddening with the values listed in Table \ref{tab:comp_objects} and respective references. The reference epoch for each object has been set to its first peak.
  } 
  \label{fig:r_band_comp}
\end{figure*}

Luminous red novae and ILRTs can be said to belong to the same broad family of red transient phenomena, and as such it can be difficult to disentangle the two observationally. For example, the [Ca~{\sc ii}] $\lambda\lambda$7291,7323 doublet, characteristic of ILRTs, is also seen in some LRNe \citep{Cai2019}. Furthermore, the forest of metal lines characteristic of LRNe can also be seen in ILRTs (e.g. NGC~300-2008-OT1; \citealt{Humphreys2011}). The ILRTs SN~2008S and NGC~300-2008-OT1 each had a detected progenitor that likely did not survive the explosive event \citep{Botticella2009, Adams2016, Valerin2024}, supporting the idea that such events are terminal.

AT~2022fnm ($\alpha=12^{\mathrm{h}}25^{\mathrm{m}}33^{\mathrm{s}}.713$, $\delta =+45^{\circ}41'12''.45$, J2000; see Fig. \ref{fig:finder}) was first discovered on 31 March 2022 (UTC dates are used throughout this paper) by the Asteroid Terrestrial-impact Last Alert System (ATLAS) survey \citep{Tonry18, SmithKW2020} and was given the internal survey designation ATLAS22kez \citep{Tonry2022}. It was classified on 2 April 2022 as a luminous blue variable by \cite{Bostroem2022} as part of the Global Supernova Project \citep{Howell2017, Howell2019}.

The region of the sky containing AT~2022fnm was well sampled by the ATLAS survey, which allowed us to constrain the epoch of the onset of the outburst to within a 3\,d window. We set the outburst epoch as MJD 59667.98 $\pm$ 1.5 based on the midpoint between the first ATLAS detection (MJD 59669.47) and the last non-detection (MJD 59666.50, with a limiting magnitude of 20.17 in the ATLAS $c$ band). 

AT~2022fnm is associated with the galaxy NGC~4389, for which there is a spread of distances reported in the literature. We adopted a luminosity distance of 10.3~Mpc, as determined from the measured redshift of 0.00239 for a Hubble constant of 69.6~km~s$^{-1}$~Mpc$^{-1}$ \citep{Bennett2014}. This results in a distance modulus of $\mu =30.1 \pm 0.6$~mag. 
This distance falls roughly between the recent Tully-Fisher distance from the Cosmicflows4 catalogue ($\mu =29.855 \pm 0.40$~mag; \citealt{Tully2013}) and the kinematic distance after correcting for local velocity fields, $\mu = 30.70 \pm 0.15$~mag, from the NASA Extragalactic Database\footnote{\url{http://ned.ipac.caltech.edu/}}(NED). We take AT~2022fnm to have a foreground extinction of $A_\mathrm{V}=0.040$~mag from \citet[from NED]{Schlafly11}, which assumes an $R_\mathrm{V}$ of 3.1 \citep{Schultz1975}; motivated by the blue spectra and unremarkable colour evolution, we assume negligible host-galaxy extinction.

This paper is organised as follows: in Section \ref{sect:obs_data} we discuss the acquisition and reduction of the optical and infrared data. In Section \ref{sect:phot_evol} we present the photometric evolution of the object and compare it to other gap transients; a full list can be found in Table \ref{tab:comp_objects}. In Section \ref{sect:spec_evol} we address the spectroscopic evolution. Section \ref{sect:disc} contains our overall discussion. The data logs, both photometric and spectroscopic, along with the details of the objects used for comparison can be found in Appendix \ref{sect:appendix_tab}.


\section{Observational data}
\label{sect:obs_data}
\subsection{Optical and infrared imaging}

We obtained optical imaging from a number of telescope and instrument configurations: Nordic Optical Telescope (NOT) + ALFOSC (as part of the NUTS2 programme\footnote{\url{https://nuts.sn.ie/}}), 
Liverpool Telescope (LT) + IO:O, Gran Telescopio Canarias (GTC) + OSIRIS, Asiago Schmidt 67/92 + Moravian, and the Las Cumbres Observatory Global Telescope (LCOGT) network of 1m telescopes. 
A detailed log is given in Table \ref{table:opt_phot}.
The images were reduced in a standard manner, comprising trimming, bias subtraction and flat-field correction. In the case of the NOT+ALFOSC and Asiago data, we performed these reductions with the {\sc foscgui} pipeline\footnote{\url{https://sngroup.oapd.inaf.it/foscgui.html}} developed by E. Cappellaro. The IO:O data were reduced using the IO:O pipeline\footnote{\url{https://telescope.livjm.ac.uk/TelInst/Pipelines/\#ioo}}. The Las Cumbres Observatory data were automatically reduced by the {\sc banzai} pipeline\footnote{\url{https://github.com/LCOGT/banzai}}. Imaging was also obtained from the ATLAS and \textit{Zwicky} Transient Facility (ZTF; \citealt{Bellm19}) surveys. 
These data are reduced automatically using bespoke pipelines \citep{Smit00,Masci2019,Magn16}. All data were taken in Sloan {\it ugriz}, Johnson-Cousins {\it BV,} and ATLAS {\it co} filter systems; the ATLAS {\it o} band is roughly equivalent to {\it r+i} and ATLAS {\it c} to {\it g+r}. 

We also obtained five epochs of NIR imaging using NOT+NOTCam; the log is given in Table \ref{table:NIR_phot}. The images were reduced using the NOTCam {\sc QUICKLOOK} reduction package\footnote{\url{http://www.not.iac.es/instruments/notcam/guide/observe.html}}. A master flat was created using bright and dim sky flats and bad pixels were masked. The wide-field camera of NOTCam suffers significant optical distortion for which we accounted using a distortion model. Sky images were created from the dithered on-source exposures, which allowed us to subtract the dark current along with the sky background. For both optical and NIR imaging, we performed point-spread-function-fitting photometry using the SuperNOva PhotometrY (SNOoPY) package\footnote{SNOoPY is a package for SN photometry using point-spread-function fitting and/or template subtraction developed by E. Cappellaro. A package description can be found at \url{http://sngroup.oapd.inaf.it/ecsnoopy.html}}.

\subsection{Optical spectroscopy}
We obtained ten epochs of long-slit optical spectra spanning +2\,d to +90\,d from a number of different telescopes; the observation log can be seen in Table \ref{table:opt_spec}.

We reduced the NOT+ALFOSC spectra using the {\sc foscgui} data reduction pipeline, whilst the Keck+LRIS spectrum was reduced using the {\sc LPipe} data-reduction pipeline \citealp{Perley2019}. The Telescopio Nazionale {\it Galileo} (TNG)+DOLORES and GTC+OSIRIS spectra were reduced manually using standard IRAF\footnote{IRAF is a software suite for the reduction and analysis of astronomical data. A description can be found at \url{https://iraf-community.github.io/}} tasks. The reduction process was similar for each instrument, with the two-dimensional (2D) spectra trimmed, bias-subtracted, and flat-fielded before being extracted to one dimension. Arc-lamp spectra were used to calibrate the wavelengths of the spectra and a master sensitivity function was created from a standard-star spectrum that was taken with an identical instrumental setup, before a telluric correction was applied. 

\section{Photometric evolution}
\label{sect:phot_evol}
\subsection{Light curve and colour}
\begin{figure}[h]
\includegraphics[width=\linewidth]{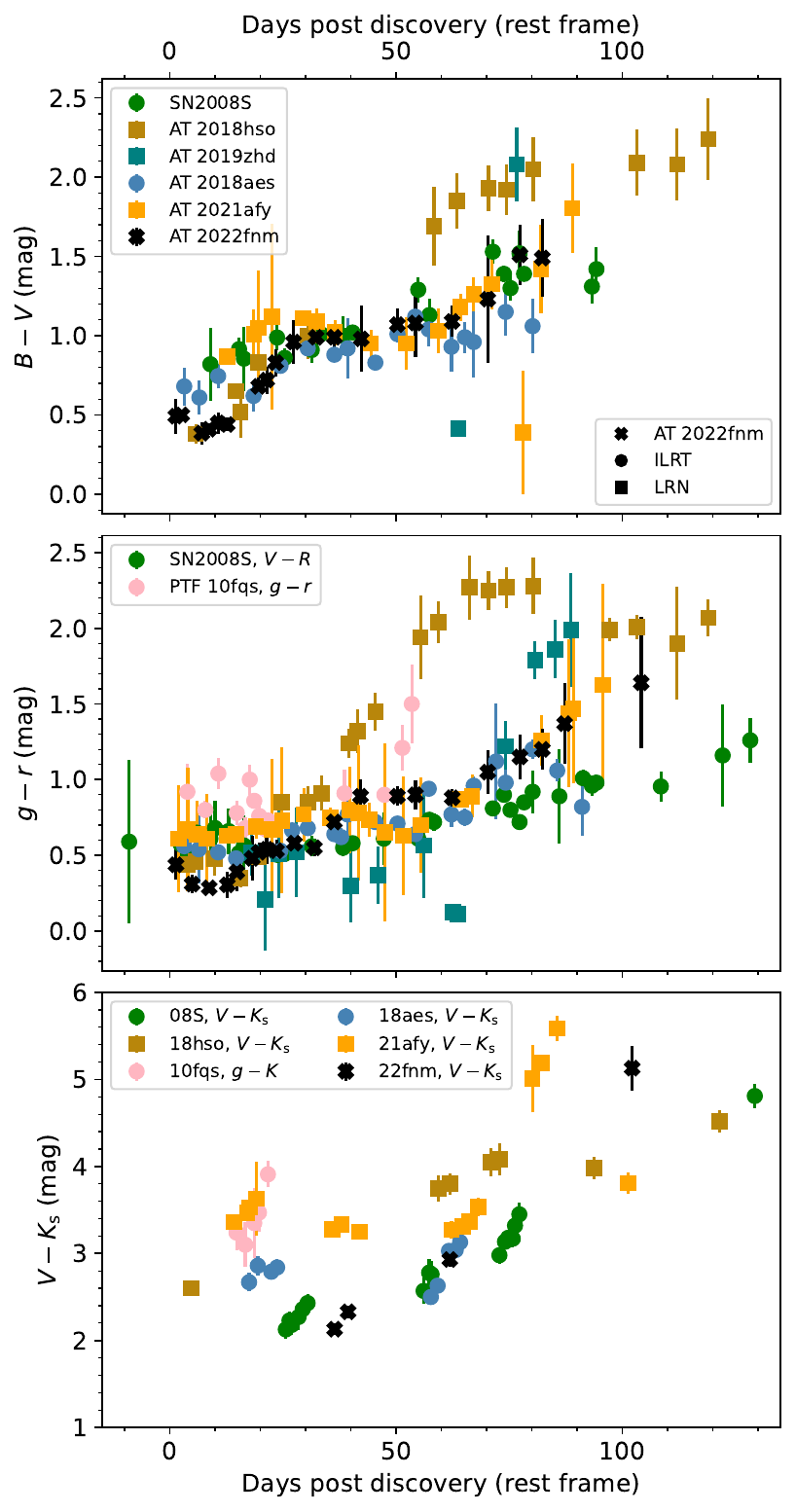}
  \caption{Colour evolution of AT~2022fnm and comparison objects. We calculated the colours with observations taken within 0.3\,d of each other except in the case of $V-K_\mathrm{s}$, for which we allowed a $\pm$3.5\,d window, owing to the modest cadence of the NIR observations.}
  \label{fig:colour}
\end{figure}

\begin{figure*}
\includegraphics[width=\linewidth]{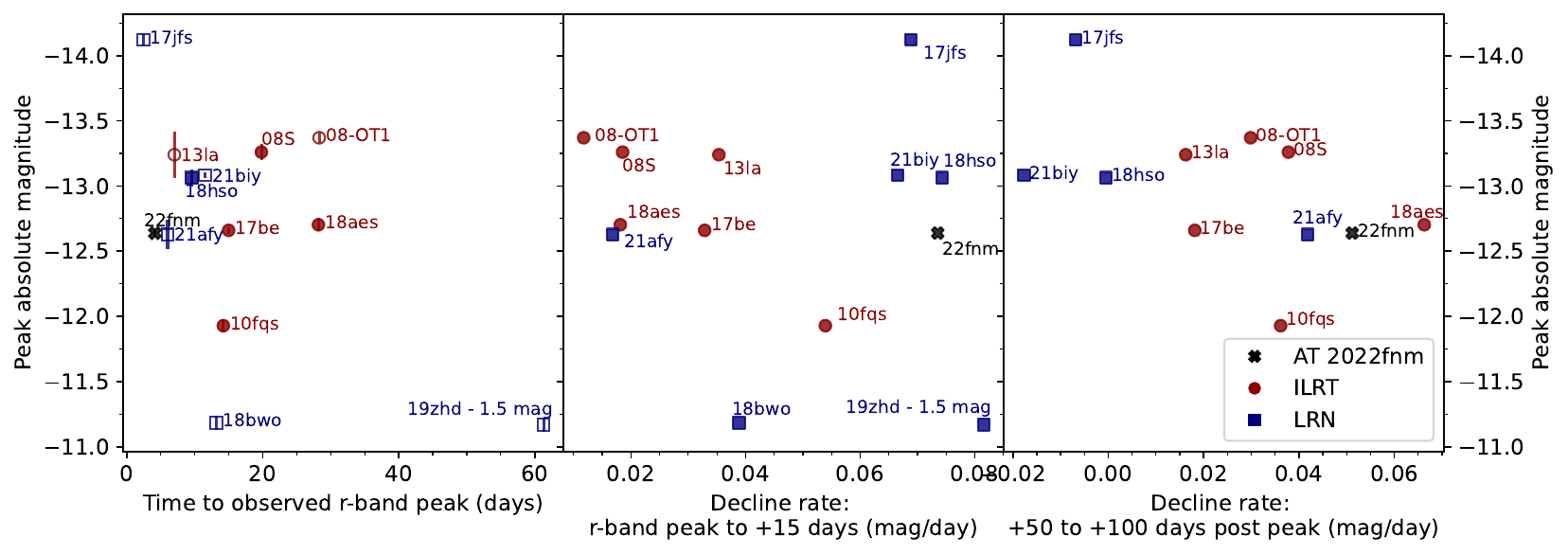}
  \caption{Rise time and decline rates plotted against the (first) $r$-band peak for AT~2022fnm and comparison objects. Rise times plotted with filled markers are calculated against explosion or outburst epoch, whilst those with unfilled markers are calculated against the detection epoch. Rise times are calculated to the peak observed magnitude, which could be more sensitive to photometric uncertainties for objects whose light curve flattens around the peak.}
  \label{fig:rise_and_decline_rates_vs_peak}
\end{figure*}

The multi-band light curve of AT~2022fnm is shown in Fig. \ref{fig:lc}. It exhibits a relatively smooth evolution in all bands, with a plateau between approximately +15 and +50\,d in the redder bands.

As the $o$-band limit is the closest to the discovery epoch, we used it to calculate the rise time by fitting it with a second-order polynomial; the resulting rise time is 5.3 $\pm$ 1.5\,d, which is relatively well constrained compared to the limited sample of ILRTs with secure rise times, and we find a peak apparent magnitude of $17.36$~mag in $o$, corresponding to an absolute magnitude of $-$12.7 $\pm$ 0.1. 
Overall, the light curve of AT~2022fnm is also reminiscent of LRNe, though it only displays a relatively short plateau rather than a second peak.

\begin{figure}
\includegraphics[width=\linewidth]{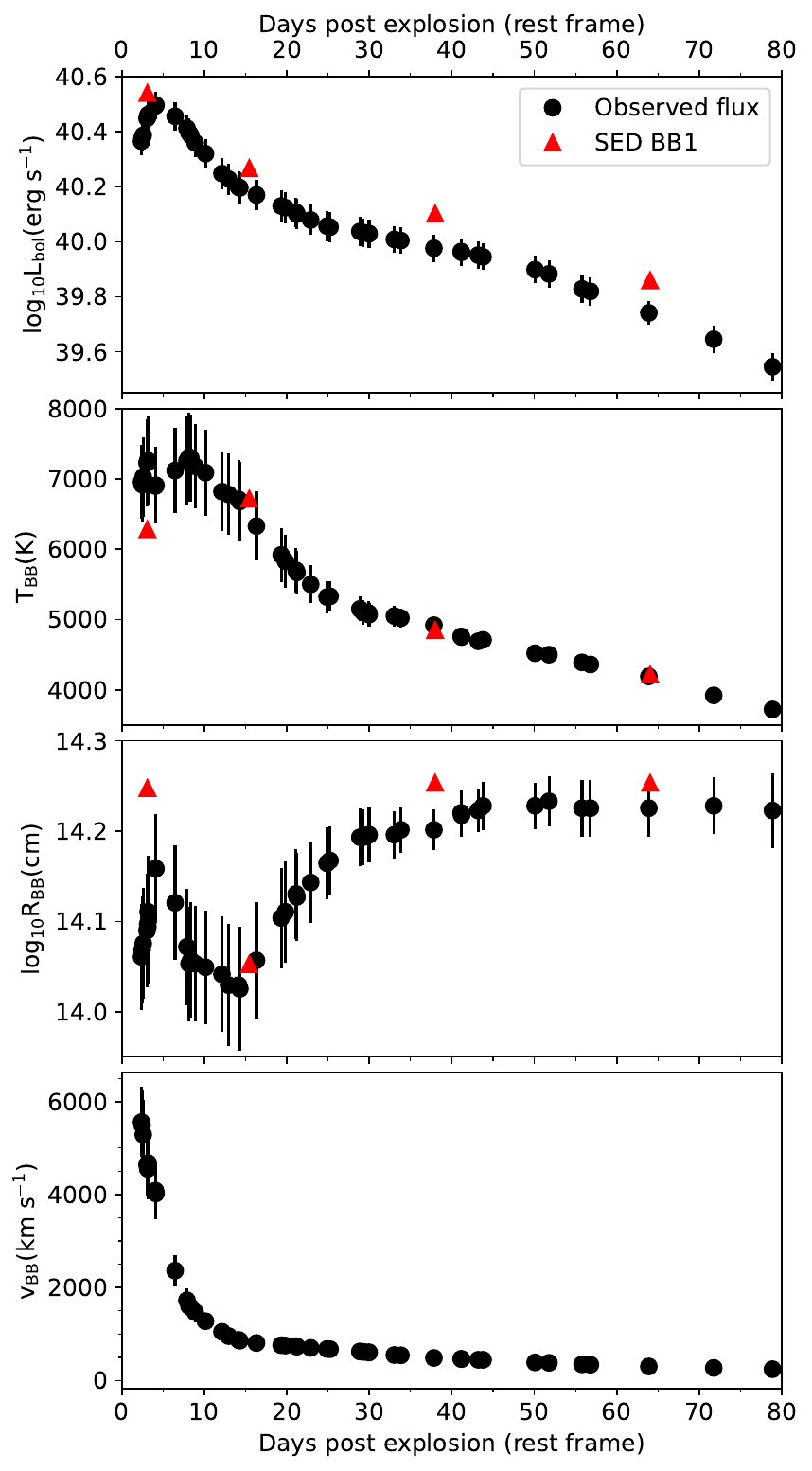}
  \caption{Pseudo-bolometric properties. {\it Top panel:} Pseudo-bolometric light curve of AT~2022fnm. {\it Second panel:} Temperature versus time of a black body that has been fitted to the pseudo-bolometric light curve. {\it Third panel:} Log of the radius versus time of a black body that has been fitted to the pseudo-bolometric light curve. {\it Bottom panel:} Expansion velocity (radius over time) versus time of a black body that has been fitted to the pseudo-bolometric light curve. Note that the radius is simply being divided by the elapsed time at each point to generate this crude measure. The data have been corrected for redshift and foreground extinction. Points corresponding to the SED fits from Fig. \ref{fig:SED} have been overplotted for comparison in the form of red triangles. The figure has been truncated at +80\,d as the light curve is dominated by numerical artefacts beyond this epoch.}
  \label{fig:pseudobolo}
\end{figure}

\begin{figure}
\includegraphics[width=\linewidth]{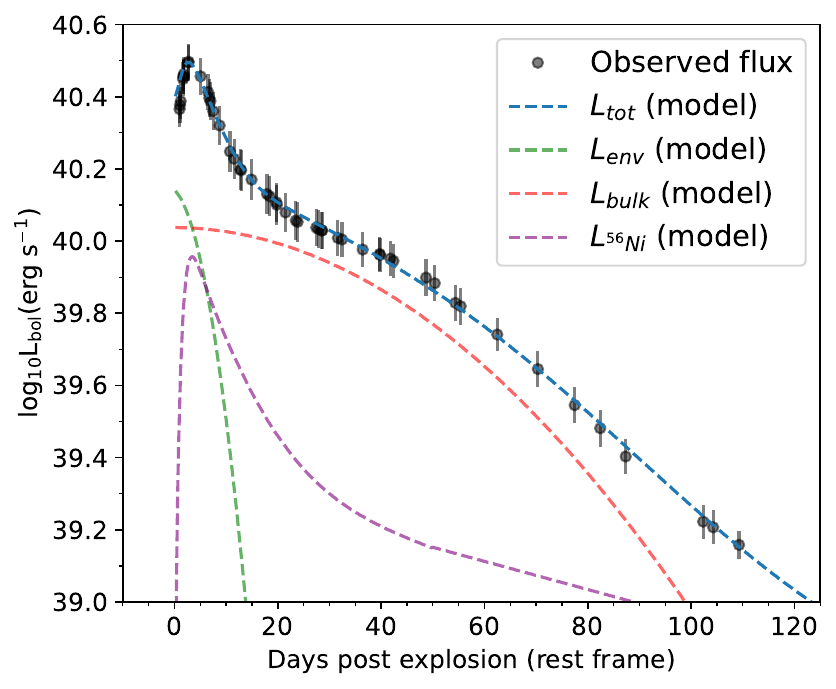} 
  \caption{Attempt to reproduce the pseudo-bolometric light curve with a simple analytical diffusion model. All three components of the model are shown along with their total, which fits the observed luminosity well. The model is somewhat degenerate, and the details of this are discussed in the text, as are the parameters. The best-fitting explosion epoch from our modelling is MJD 59669.38, which is consistent with our adopted outburst epoch (MJD 59667.98 $\pm$ 1.5).} 
  \label{fig:modelling}
\end{figure}

We compare the $r$-band light curve of AT~2022fnm with  ILRTs and LRNe in Fig. \ref{fig:r_band_comp} (references in \ref{tab:comp_objects}). For the first 100\,d, the light-curve evolution of SN~2008S does not differ markedly from that of AT~2022fnm, but the rate of decline is slower and it does not display the same clear peak and plateau that is visible in the AT~2022fnm light curve.
AT~2018aes \citep{Cai2021} has a similar peak to AT~2022fnm, but its behaviour over the first 40\,d after peak does not resemble AT~2022fnm at all, instead showing much slower evolution, but after +40\,d their decline rates are quite similar. PTF~10fqs, albeit fainter than AT~2022fnm, shows a similar evolution, though it has a clear plateau that lasts until about +70\,d.

In Fig. \ref{fig:colour} we show the colour evolution of AT~2022fnm in $B-V$, $g-r$, and $V-K_\mathrm{s}$; all display a clear redward progression with time following a brief blueward excursion in the first ten days for $B-V$ and $g-r$ (we do not have $V-K_\mathrm{s}$ data covering this period). AT~2021afy, a LRN, shows qualitatively similar behaviour to AT~2022fnm in $B-V$, with a particularly close match in evolution from about +40\,d. On the other hand, SN~2008S, the prototypical ILRT, also shows similar evolution in this colour. The limited data for AT~2022fnm in $V-K_\mathrm{s}$ make comparison more difficult, but in general, we see a redward movement in both ILRTs and LRNe, though the rate varies from object to object. 
SN~2008S, despite its slightly less steep redward movement, is qualitatively the best match to AT~2022fnm.

In Fig. \ref{fig:rise_and_decline_rates_vs_peak} we show both the rise time of AT~2022fnm and our comparison objects and their decline rates over the first 15\,d and over the period between +50\,d and +100\,d (where applicable). The purpose of this plot was to determine where AT~2022fnm fell in the parameter space and whether or not it showed a particular kinship with either LRNe or ILRTs. At least in the case of our limited sample, it seems that LRNe occupy a larger portion of the parameter space both in terms of rise times and decline rates, whilst the ILRTs are clustered together. This is consistent with the findings of \citealt{Cai2022a}. 
AT~2022fnm occupies a region consistent with both LRNe and ILRTs.

\subsection{Pseudo-bolometric light curve and spectral energy distribution evolution}

In Fig. \ref{fig:pseudobolo} we show the pseudo-bolometric light curve of AT~2022fnm along with evolution of the corresponding black-body temperature and radius. The light curve was computed based on our $uBgcVroizJHK_\mathrm{s}$ data using the {\sc superbol} package.\footnote{\url{https://github.com/mnicholl/superbol}} \citep{Nicholl2018}. 
Owing to a gap in the data coverage, we only show the first 80\,d (cf. Fig. \ref{fig:lc}). After this point the interpolation between epochs and extrapolation across different filters become unreliable, and the pseudo-bolometric light curve is dominated by numerical artefacts. We see that the black-body radius first decreases (around +5\,d), before increasing again at around +15\,d. Hydrogen recombination could increase the transparency, but then another energy source would be required to stop and reverse this behaviour. As such, a source of ionising photons is required to push the photosphere back out; this will be addressed in Section \ref{sect:disc}. The photospheric velocities evident in the bottom panel of Fig. \ref{fig:pseudobolo} are 4,000 \kms\ and higher in the first week, whilst the spectra do not suggest velocities this high. However, the spectral lines in emission form farther out in an optically thin region of the circumstellar material (CSM), whereas the velocities derived from the black-body fitting reflect an optically thick continuum that lies at a smaller radius.

\begin{table*}
\caption{Analytical diffusion model parameters.}
\centering
\begin{tabular}{lllllllllll}
\hline
\hline
$M_\mathrm{env}$ & $R_\mathrm{env}$ & $v_\mathrm{env}$ & $E_\mathrm{env}$ & $M_\mathrm{bulk}$ & $R_\mathrm{bulk}$ & $v_\mathrm{bulk}$ & $E_\mathrm{bulk}$ & $E_\mathrm{tot}$ & $M_\mathrm{Ni}$ & $t_\mathrm{diff,^{56}Ni}$  \\
\msun & cm & \kms & erg & \msun & cm & \kms  & erg & erg & \msun & days \\
\hline
0.15 & 10$^{14}$ & 9000 & 4x10$^{46}$ & 4 & 2x10$^{13}$ & 5000 & 4.2x10$^{48}$ & 4.24x10$^{48}$ & 1.6x10$^{-4}$ & 2.3 \\
\hline
\end{tabular}
\label{table:model_params}
\end{table*}

We attempted to reproduce the pseudo-bolometric light curve with a simple analytical diffusion model that accounts for the luminosity of the cooling ejecta and the \nic~decay process. Details will be presented in \cite{Valerin2024}, but the fundamental design is based on \citet{Chatzopoulos2012} and \citet{Arnett1980, Arnett1982}. The resulting model light curve for AT~2022fnm is shown in Fig. \ref{fig:modelling}. 
To mimic both the sharp first peak as well as the subsequent slow decline, we divided the ejecta into two regions. The outer layers (which can be identified with the stellar envelope) are faster and less dense, and therefore release their internal energy over a short timescale. On the other hand, the inner parts of the ejecta contain the bulk of the mass and are slower, resulting in a longer photon diffusion time. Hydrogen recombination, though not directly treated, is accounted for in a rough manner by means of the so-called diffusion time in the bulk material. The light curve is well reproduced by the model, with both the peak and the decline captured.

We note that the mass and the velocity of the ejecta are coupled, which means that an equivalent model could be contrived with a higher mass and velocity or, equally, with a lower mass and velocity, so we cannot use this model to constrain the energy of the outburst reliably.
Nevertheless, we were able to reproduce the pseudo-bolometric light curve with a plausible set of parameters. The model considers the integrated radiated energy of the bulk material and the envelope separately, but combined they give a total integrated radiated energy of $E_\mathrm{tot} = 4.24 \times 10^{48}$~erg. 
The best-fitting explosion epoch from our modelling is MJD 59669.38, which agrees within the uncertainties with our adopted outburst epoch (MJD 59667.98 $\pm$ 1.5). Overall, the model is consistent with a weak eruption or explosion of a star of relatively low mass. The full set of model parameters can be found in Table \ref{table:model_params}. More sophisticated modelling may be used in the future to provide more robust explosion parameters.

\begin{figure}
\includegraphics[width=\linewidth]{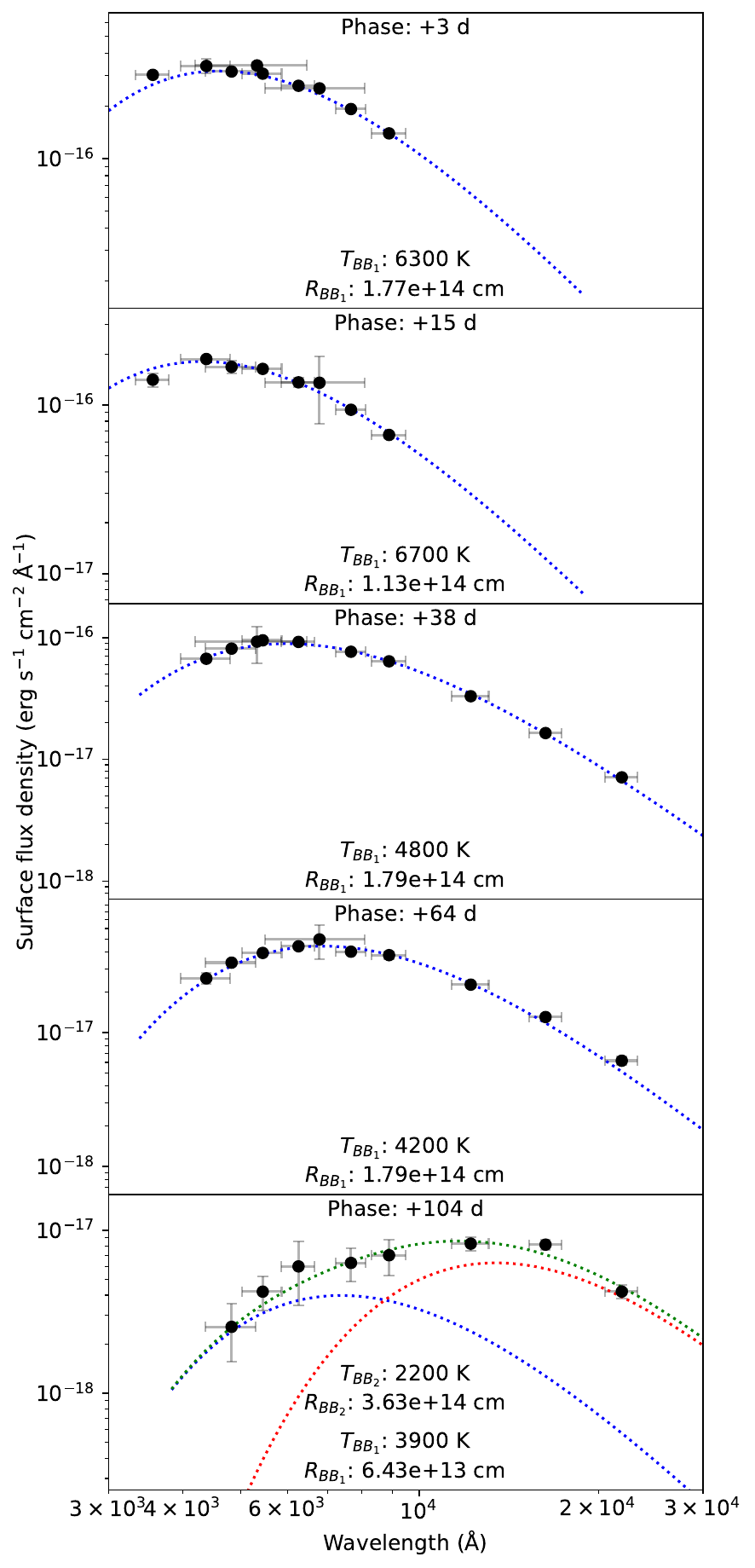}
  \caption{Sequence of SED fits for AT~2022fnm. At late times a second black-body component (red) is required, and the sum of the two components is shown in green.}
  \label{fig:SED}
\end{figure}

In Fig. \ref{fig:SED} we show a sequence of spectral energy distribution (SED) fits for AT~2022fnm. It can be seen that at late times the SED is best fit by two black bodies, owing to the development of a NIR excess. The NIR photometric data are sparse, though it does seem that the parameters of the second black body stay fairly constant between +104\,d and +120\,d. 

\section{Spectral evolution}
\label{sect:spec_evol}
\begin{figure*}
\includegraphics[width=0.97\linewidth]{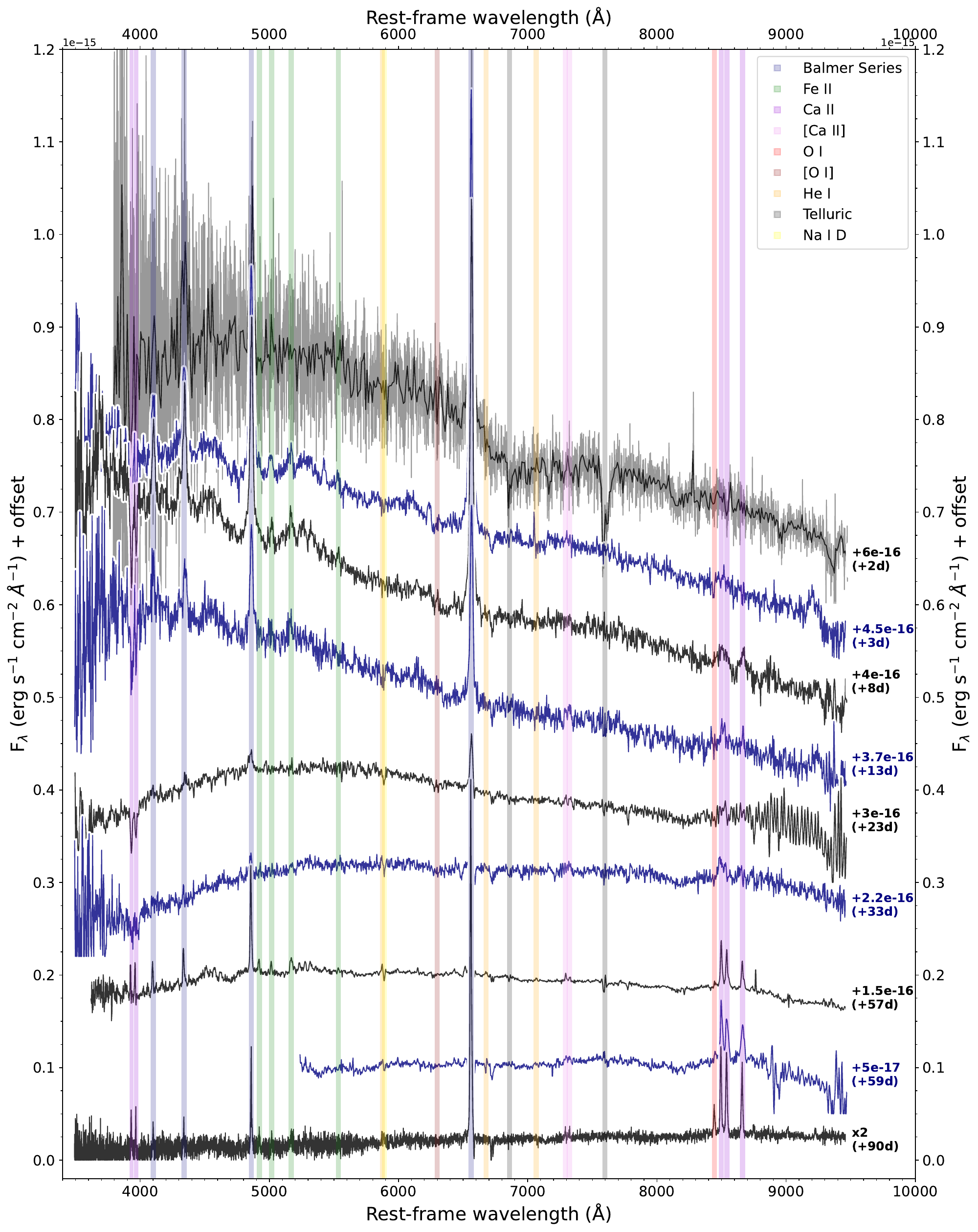}
  \caption{Optical spectral sequence of AT~2022fnm. The spectra have been calibrated against the photometry. The offsets applied to each spectrum are shown in the plot, and the +90\,d spectrum has been multiplied by a factor of 2 for clarity. Negative fluxes have been set to zero. The +2\,d spectrum is plotted in grey with a version binned to 10~$\AA$ overplotted in black to compensate for the noise of the spectrum. The absorption features seen at approximately 6725\,\AA\ and 8200\,\AA\ are artefacts.}
  \label{fig:spectra}
\end{figure*}

The spectral sequence for AT~2022fnm is shown in Fig. \ref{fig:spectra}. 
It initially displays a strong blue continuum that weakens with time, and in the later epochs the continuum clearly  becomes red. There is a large decrease in relative luminosity bluewards of H$\beta$ between the +13\,d and +23\,d spectra. The development of such a red continuum is common to both ILRTs and LRNe.

\begin{figure*}
\includegraphics[width=0.97\linewidth]{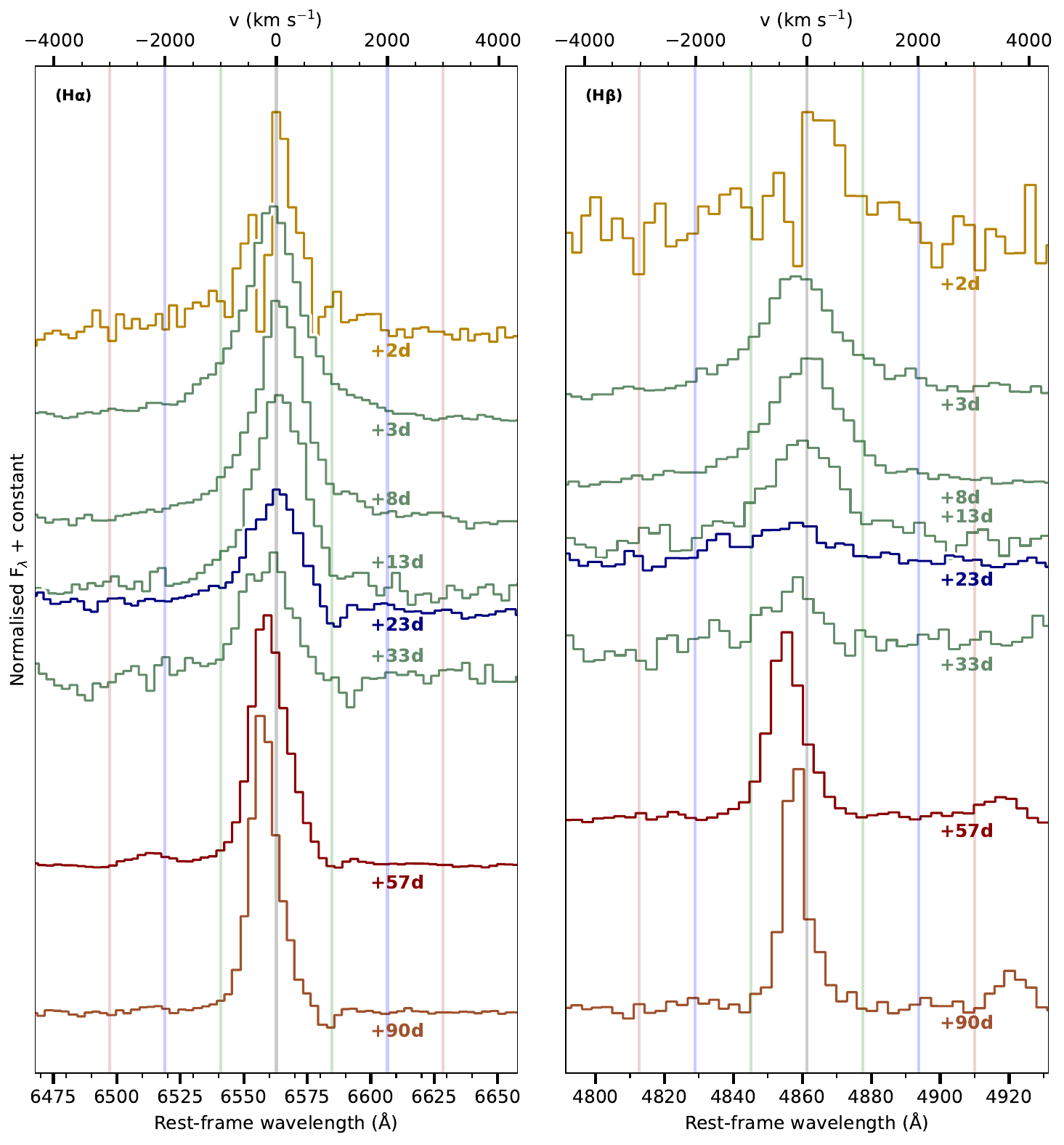}
  \caption{Evolution of H$\alpha$ and H$\beta$ in AT~2022fnm over time. All spectra have been binned to a common pixel size and normalised against the peak of H$\alpha$ in the left panel and H$\beta$ in the right. Only spectra with both H$\alpha$ and H$\beta$ coverage are shown. Vertical lines marking $\pm 1000$, 2000, and 3000~\kms\ have been added to guide the eye.}
  \label{fig:Halpha_Hbeta_evol}
\end{figure*}

Balmer lines dominate the spectra throughout the evolution of AT~2022fnm. Balmer lines are seen in both ILRTs and LRNe, though at late times the lines typically decrease in prominence significantly in LRNe. 
The complete evolution of H$\alpha$ and H$\beta$ is plotted in Fig. \ref{fig:Halpha_Hbeta_evol} with the spectra having been binned to a common sampling. 

It can be seen in Fig. \ref{fig:Halpha_Hbeta_evol} that the H$\alpha$ emission feature narrows with time and that H$\beta$ shows similar behaviour. An absorption line is present bluewards of H$\alpha$ at 6485\,\AA\ in the +33\,d spectrum, which we associate with H$\alpha$. A similar absorption is present bluewards of H$\beta$ at 4802\,\AA. 
Both have velocities of approximately $-3500$~\kms. In the +33\,d spectrum, two absorption features can be seen at 6574\,\AA\ and 6592\,\AA, just redwards of H$\alpha$, and if they are associated with it, then they have velocities of approximately $+500$~\kms\ and $+1300$~\kms, respectively.

There is a noticeable H$\alpha$ absorption trough in the +2\,d spectrum at approximately 6559\,\AA, corresponding to a velocity of around $-150$~\kms\ and with a full width at half maximum of approximately 240~\kms. This feature is not present in the +3\,d or later spectra. If this feature is real, then it must result from a thin shell of material in the immediate vicinity of the outburst that was quickly overrun. 

\begin{figure*}
\includegraphics[width=0.97\linewidth]{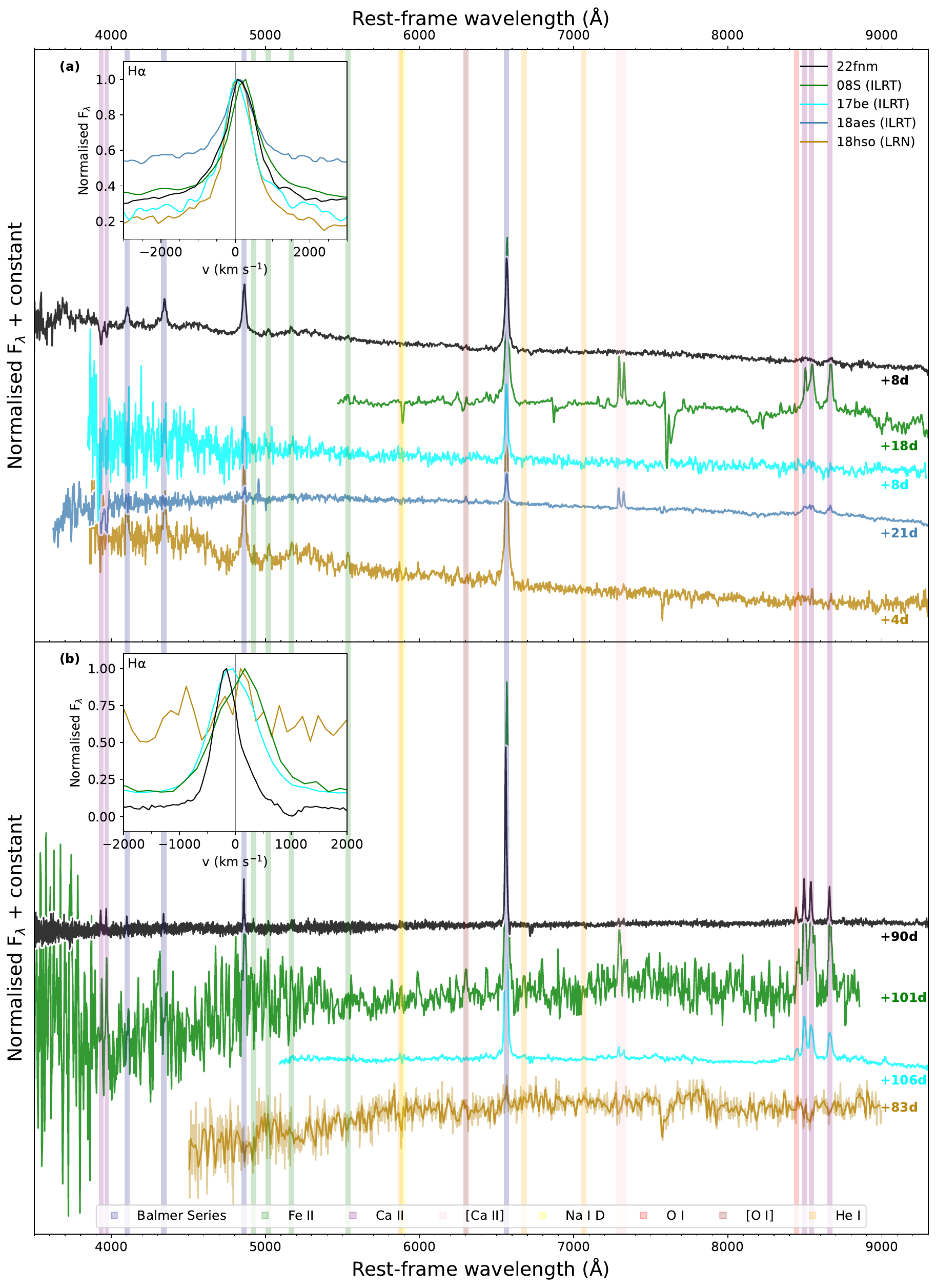}
  \caption{Comparison of an early spectrum and late spectrum of AT~2022fnm with other objects. The spectra have been normalised against the sum of the flux densities except in the case of the insets, which have been normalised against the peak of H$\alpha$. The phases are calculated relative to the epoch of the onset of the outburst. The upper-left inset shows H$\alpha$. The +83\,d AT~2018hso spectrum has been trimmed to 4500--9000\,\AA\ and binned to 10\,$\AA$, owing to noise.}
  \label{fig:spec_comp}
\end{figure*}

To assess the evolution of the Balmer lines, we fitted a Lorentzian profile at each epoch to adequately capture the electron-scattering wings. We find that over the first 15\,d after the onset of the outburst, the full width at half maximum velocity stays at approximately 625--750~\kms\ but drops monotonically to around 400~\kms\ at +90\,d. 
In Fig. \ref{fig:spec_comp} we compare one early (+8\,d) and one late spectrum (+90\,d) of AT~2022fnm to spectra of other objects. In the upper-left inset of each panel, we show a zoom-in of the H$\alpha$ region for each object. In the upper-left inset of panel (a), the H$\alpha$ profile of AT~2022fnm resembles that of AT~2017be and SN~2008S in both shape and velocity, though the peak of the latter appears to be slightly redshifted.
In the upper-left inset of panel (b), H$\alpha$ appears to be slightly blueshifted at late times in the case of AT~2022fnm and AT~2017be, but slightly redshifted in the case of SN~2008S and perhaps AT~2018hso, though the signal-to-noise ratio of the line is not particularly good for that object. 
At early times all of the objects (AT~2022fnm, SN~2008S, AT~2017be, AT~2018aes, and AT~2018hso) are dominated by the Balmer lines, as is typical in ILRTs and LRNe. The overall spectral evolution of AT~2022fnm resembles that of the ILRTs AT~2017be and AT~2018aes. The LRN AT~2018hso resembles AT~2022fnm at early times, but at late times it displays a red continuum and there is little resemblance between the two objects in terms of spectral lines, firmly placing AT~2022fnm in ILRT territory, as far as its spectra are concerned. 

Moving on from the Balmer lines to consider oxygen, we note that there is the narrow absorption at around 7770\,\AA, which is visible in the +33\,d spectrum as well as the preceding spectra. If this line is O~{\sc I} $\lambda$7774, then it is unusual to see it in absorption; it does not appear to be visible in galactic emission in the 2D spectra, so it is unclear whether this is a result of subtraction. Typically, the O~{\sc I} $\lambda$7774 line is seen in emission in airglow spectra. O~{\sc I} emission at 8446\,\AA\ is apparent by +57\,d, growing to much greater prominence by +90\,d. This O~{\sc I} emission is not pronounced in SN~2008S or AT~2017be at early times, but is clear at later times in these objects, as can be seen in Fig. \ref{fig:spec_comp}.

It is difficult to confirm the presence of Fe~II above the noise in the +2\,d spectrum, but it is clearly present from the +3\,d spectrum all the way up through the +90\,d spectrum. In particular, there are emission lines at 4924\,\AA, 5018\,\AA, and 5169\,\AA, as well as a forest of emission lines between approximately 4490\,\AA\ and 4620\,\AA.

Turning to calcium, initially weak, we note that the Ca~{\sc ii} $\lambda$8498,8542,8662 NIR triplet becomes apparent within the first two weeks and is extremely pronounced at late times. The triplet is also weak at early times in AT~2017be (Fig. \ref{fig:spec_comp}), but, as in AT~2022fnm, it is very prominent at later times. The [Ca~{\sc ii}] $\lambda\lambda$7291,7323 doublet, characteristic of ILRTs, though present from the +13\,d spectrum onwards, is unusually weak for an ILRT. 
Like AT~2022fnm, the late-time spectrum of the ILRT AT~2017be does not show particularly prominent emission in the [Ca~{\sc ii}] $\lambda\lambda$7291,7323 doublet (Fig. \ref{fig:spec_comp}). The doublet, a key signature of ILRTs, is not typically seen in LRNe, although it was observed in AT~2018hso as a very weak emission feature \citep{Cai2019}. 
The Ca~H $\lambda$3969 and Ca~K $\lambda$3934 lines are initially very prominent in absorption, but by +57\,d they are instead in emission. The change from Ca~H and K absorption to emission can be seen in Fig. \ref{fig:spec_comp} where Ca~H and K emission is also seen in the ILRTs SN~2008S and AT~2018aes as well as in the LRN AT~2018hso.

The late-time spectra are reminiscent of interacting SNe, although we note that many CSM-dominated transients appear similar at these phases. We see no evidence for spectral features associated with nucleosynthesis (e.g. [O~{\sc ii}] $\lambda\lambda$6300,6364), although these can be suppressed in interacting core-collapse SNe due to higher densities. \ion{O}{i} $\lambda$8446 is present, and it is possible that this is due to Bowen fluorescence \citep{Bowen47}. In this case, the Lyman-$\beta$ photons required presumably arise from CSM interaction. 
Given that the Ca~{\sc ii} $\lambda$8498,8542,8662 NIR triplet and that lines due to Fe~{\sc ii} are present, whilst the [Ca~{\sc ii}] $\lambda\lambda$7291,7323 doublet is very weak (Fig. \ref{fig:spectra}) and [Fe~{\sc ii}] lines are absent, suggests that the density remained high \citep{Kozma1998, Dessart2011}.

\section{Discussion}
\label{sect:disc}
The similarity of the light curve of the ILRT AT~2022fnm with light curves of LRNe reflects that in both cases a comparable amount of energy is injected into a comparable amount of hydrogen-rich material. The object displays a clear NIR excess (apparent by approximately +100\,d), typical of ILRTs \citep{Cai2021} and also seen in a number of LRNe \citep{Pastorello2019c, Stritzinger2020a}. On the other hand, the object does not display the double-peaked light curve associated with LRNe, though this feature is not particularly prominent in some LRNe (e.g. AT~2018hso; \citealt{Cai2019}). Nor does AT~2022fnm display the overwhelming forest of metal lines in absorption along with the weakening of the Balmer lines typically seen at intermediate times when LRNe have their second peak, or the molecular lines associated with LRNe at late times. In certain respects, AT~2022fnm resembles AT~2018hso, though AT~2018hso is a LRN with a double-peaked light curve.

Unlike LRNe, which display a second peak in photospheric radius after the maximum, the photospheric radii of ILRTs typically undergo a monotonic decrease \citep{Botticella2009, Cai2019, Cai2021}. AT~2022fnm, on the other hand, displays aberrant behaviour in this regard, which precludes the use of this metric as a tidy method of separating LRNe and ILRTs. In Fig. \ref{fig:pseudobolo} over the first 20\,d we see the photosphere moving inwards and then outwards again. The decrease in radius could potentially be explained by increasing transparency due to hydrogen recombination, which is indeed seen at around 6000\,K in SN IIP models, for example \citep{Kasen2009}.

After the decrease in radius, a source of ionising photons would then be required to push the photosphere outwards. The spectral evolution provides some clues: the continuum towards the blue end weakens between +13\,d and +23\,d, which is potentially indicative of increased opacity as the bluer wavelengths experience greater scattering. Furthermore, the beginning of the increase in radius at around +15\,d also corresponds to the beginning of the plateau seen in the light curve. Taken together, these observations could na{\"\i}vely be seen to suggest increased interaction, with the outer layers of ejecta hitting the CSM and leading to the creation of photons and hence ionisation, increasing the opacity. Alternatively, we may be seeing a scenario similar to the increase in opacity routinely observed in SNe~IIP, where the shock ionises the outer envelope, leading to a decrease in transparency. The radius of AT~2022fnm shows an overall variation of approximately a factor of 1.6, whilst the low-luminosity classical IIP SNe SN~2005cs and SN~2009md display variation of approximately a factor of 2.5 \citep{Jager2020}. However, in contrast to AT~2022fnm, the early evolution of SN~2005cs shows a monotonic increase in radius over time (unfortunately, no early radius measurements are available for SN~2009md; \citealp{Jager2020}).

\begin{figure}
\includegraphics[width=0.97\linewidth]{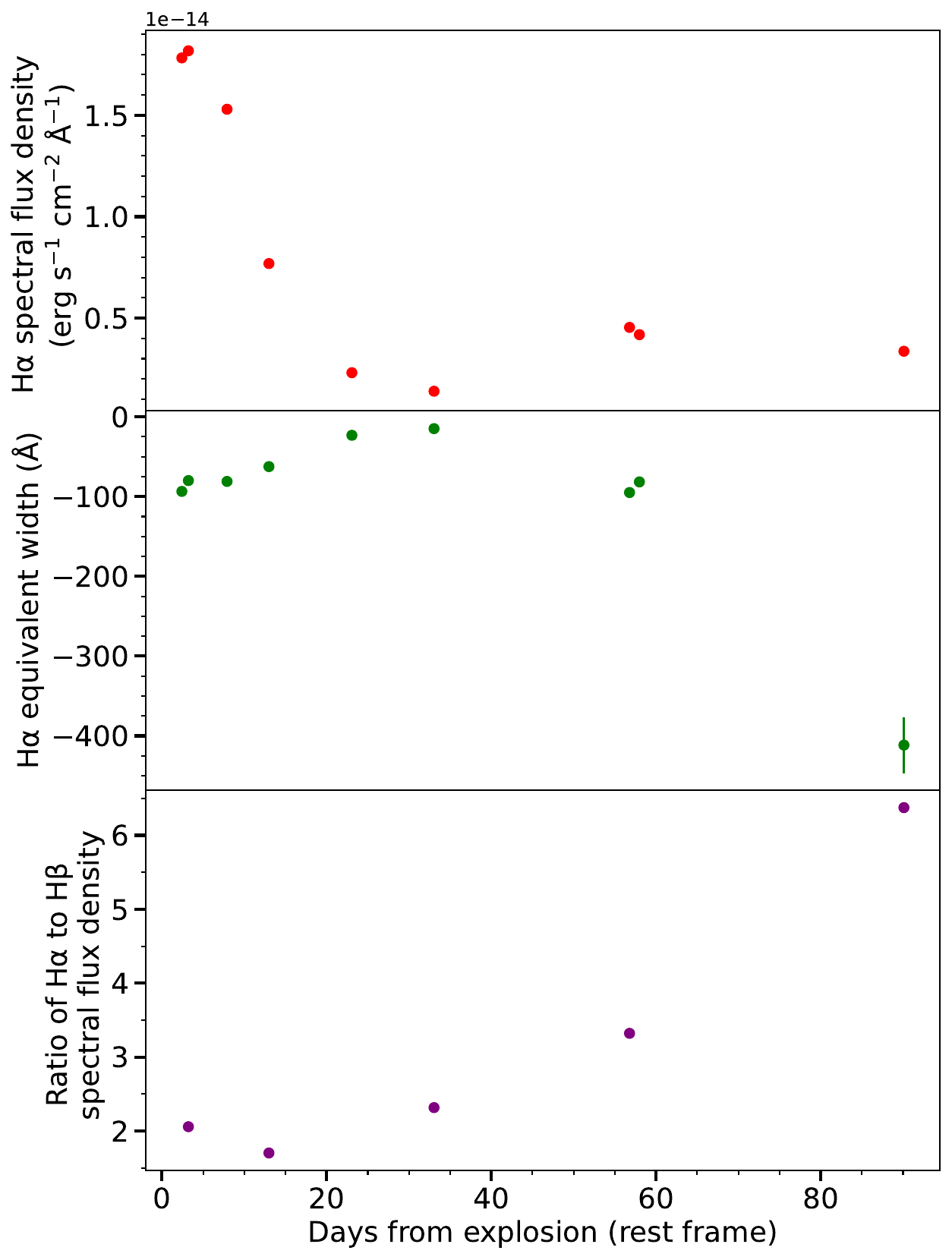}
  \caption{Evolution of the spectral flux density and the equivalent width of H$\alpha$. Top panel: Evolution of the spectral flux density of H$\alpha$ over time. Middle panel: Evolution of the equivalent width of H$\alpha$ over time. Bottom panel: Evolution of the ratio of H$\alpha$ to H$\beta$ spectral flux density over time.}
  \label{fig:eqw}
\end{figure}

The increase in radius in AT~2022fnm from around +15\,d is puzzling given that no temperature rise corresponding to the proposed ionisation is seen, though the decline rate of the luminosity does slow at this epoch. Additionally, the equivalent width of H$\alpha$  (plotted in Fig. \ref{fig:eqw}) is decreasing at this point. Similar, though less marked, behaviour was seen in the case of the ILRT SNhunt120 \citep{Stritzinger2020b}, whose photospheric radius also dipped early  before climbing again. 

It is unclear whether the requirement of a second black-body component after around +100\,d is due to the presence of multiple ejecta components or the presence of pre-existing dust. If there is pre-existing dust, it must not exist in the immediate vicinity of AT 2023fnm or it would be  quickly overrun by the ejecta. The second black-body component has a temperature of $2200 \pm 350$~K, consistent with the sublimation temperature for amorphous carbon dust of approximately 1900~K \citep{Landt2023}. Unfortunately, the point at which a second black-body component becomes necessary is approximately coincident with the epoch of our final spectrum, and the lack of later spectra makes its difficult to make any further determination.

Spectroscopically, AT~2022fnm more closely resembles an ILRT than a LRN. Whilst the development of a red continuum is common to both LRNe and ILRTs, the presence of the [Ca~{\sc ii}] $\lambda\lambda$7291,7323 doublet is a key signature of ILRTs, though, as noted, it has also been seen in the LRN AT~2018hso. Ca H and K emission is visible in the +57\,d and +90\,d spectra, and in earlier spectra it is visible in absorption. However, these lines are also seen in emission in some LRNe and ILRTs (e.g. AT~2018hso and SN~2008S). The Ca~{\sc ii} $\lambda$8498,8542,8662 NIR triplet is also prominent in both ILRTs and LRNe, as seen in Fig. \ref{fig:spec_comp}. Nevertheless, AT~2022fnm does not develop the forest of metal lines typically seen in absorption in LRNe, nor does it develop the strong molecular lines associated with LRNe at late times. This highlights the need for late-time spectra to disentangle these two classes of objects.

The absorption lines bluewards of H$\alpha$ and H$\beta$ (at 6485\,\AA\ and 4802\,\AA, respectively), which we identify as components of these lines, have a velocity of about $-3500$~\kms. This is likely associated with the ejecta as the photosphere recedes in mass coordinates, revealing further material. Indeed, this recession in mass coordinates is apparent in Fig. \ref{fig:pseudobolo} from around +30\,d, when the photospheric radius stays roughly constant. There are also further absorption features redwards of H$\alpha$ at 6574\,\AA\ and 6592\,\AA, with velocities of 500~\kms\ and 1300~\kms, respectively. 

AT~2022fnm can be said to be an ambiguous object in that it shows features consistent with LRNe and ILRTs, straddling the phase space covered by both categories of objects. AT~2022fnm is not alone in this respect, with certain other objects, such as AT~2018hso, sharing features of both the LRN and ILRT classifications \citep{Cai2019}. The simple analytical modelling we performed shows that the object is consistent with a weak eruption or explosion of a star of relatively low mass, though, as mentioned, the parameters of the model are highly degenerate. The presence of pre-existing dust, as evidenced by the NIR excess, is consistent with an ILRT; however, both pre-existing and newly formed dust can also be found in core-collapse SNe, such as SN~1987A \citep{Roche1989, Lucy1989} and the type IIP SNe SN~2002hh \citep{Meikle2006} and SN~2004et \citep{Kotak2009}, as well as in stellar-merger scenarios \citep{Bond2003, Pastorello2019c}, so it cannot be used to discriminate between these possibilities. Nevertheless, the preponderance of evidence supports the classification of AT~2022fnm as an ILRT, and the characteristics of the object are consistent with a weak stellar explosion or eruption. We expect the next generation of surveys to yield hundreds of LRNe and ILRTs, and suggest that a combination of relatively high-cadence optical and NIR photometry and spectroscopy will be needed if we wish to attribute various processes associated with the evolution of single and massive binary systems.

\begin{acknowledgements}
We thank the anonymous referee for carefully reading the manuscript. S. Moran acknowledges support from the Magnus Ehrnrooth Foundation and the Vilho, Yrj\"{o} and Kalle V\"{a}is\"{a}l\"{a} Foundation. 
RK acknowledges support from the Research Council of Finland (340613). 
MF is supported by a Royal Society -- Science Foundation Ireland University Research Fellowship. 
AP, EC and NER and AR acknowledge support of the PRIN-INAF 2022 project ``Shedding light on the nature of gap transients: from the observations to the models''.
Y.-Z. Cai is supported by the National Natural Science Foundation of China (NSFC, Grant 12303054), the Yunnan Fundamental Research Projects (Grant 202401AU070063) and the International Centre of Supernovae, Yunnan Key Laboratory (Grant 202302AN360001).
S. Mattila acknowledges support from the Academy of Finland project 350458. 
CPG acknowledges financial support from the Secretary of Universities and Research (Government of Catalonia) and by the Horizon 2020 Research and Innovation Programme of the European Union under the Marie Sk\l{}odowska-Curie and the Beatriu de Pin\'os 2021 BP 00168 programme, from the Spanish Ministerio de Ciencia e Innovaci\'on (MCIN) and the Agencia Estatal de Investigaci\'on (AEI) 10.13039/501100011033 under the PID2020-115253GA-I00 HOSTFLOWS project, and the programme Unidad de Excelencia Mar\'ia de Maeztu CEX2020-001058-M. 
NER acknowledges partial support from MIUR, PRIN 2017 (grant 20179ZF5KS) and from the Spanish MICINN grant PID2019-108709GB-I00 and FEDER funds. 
PL wishes to thank the Swedish Research Council for support. 
AVF's group at UC Berkeley has received financial assistance from the Christopher R. Redlich Fund, Gary and Cynthia Bengier, Clark and Sharon Winslow, Alan Eustace (W.Z. is a Bengier-Winslow-Eustace Specialist in Astronomy), William Draper, Timothy and Melissa Draper, Briggs and Kathleen Wood, Sanford Robertson (T.G.B. is a Draper-Wood-Robertson Specialist in Astronomy), and  numerous other donors; additional support was provide by NASA/HST grant AR-14295 from the Space Telescope Science Institute (STScI), which is operated by the Association of Universities for Research in Astronomy (AURA), Inc., under NASA contract NAS5-26555. 
This work is supported by the National Natural Science Foundation of China (NSFC grants 12033003 and 11633002), the Scholar Programme of Beijing Academy of Science and Technology (DZ:BS202002) and the Tencent Xplorer Prize.

Based in part on observations made with the Nordic Optical Telescope, owned in collaboration by the University of Turku and Aarhus University, and operated jointly by Aarhus University, the University of Turku and the University of Oslo, representing Denmark, Finland and Norway, the University of Iceland and Stockholm University at the Observatorio del Roque de los Muchachos, La Palma, Spain, of the Instituto de Astrofisica de Canarias. The data presented here were obtained in part with ALFOSC, which is provided by the Instituto de Astrofisica de Andalucia (IAA) under a joint agreement with the University of Copenhagen and NOT. 
Based in part on observations made with the Gran Telescopio Canarias (GTC), installed in the Spanish Observatorio del Roque de los Muchachos of the Instituto de Astrofísica de Canarias, in the island of La Palma. 
The W. M. Keck Observatory is operated as a scientific partnership among the California Institute of Technology, the University of California and NASA; the observatory was made possible by the generous financial support of the W. M. Keck Foundation. We thank WeiKang Zheng and Yi Yang for assistance with the Keck observations.
This work has made use of data from the Asteroid Terrestrial-impact Last Alert System (ATLAS) project. ATLAS is primarily funded to search for near-Earth objects through NASA grants NN12AR55G, 80NSSC18K0284, and 80NSSC18K1575; byproducts of the NEO search include images and catalogues from the survey area. The ATLAS science products have been made possible through the contributions of the University of Hawaii Institute for Astronomy, the Queen's University Belfast, the Space Telescope Science Institute, and the South African Astronomical Observatory. 
This work makes use of observations from the LCOGT network. 
We acknowledge ESA Gaia, DPAC, and the Photometric Science Alerts Team.
\end{acknowledgements}

%
%
\bibliographystyle{aa} 
\bibliography{bibliography}

\begin{appendix}
\section{Tables}
\label{sect:appendix_tab}
\begin{table*}
\caption{Comparison objects.}
\label{tab:comp_objects}
\centering
\begin{tabular}{lllllllp{2.5cm}}
\hline\hline
    Name & $z$ & Type &  Peak ($r$ or $R$) & $A_\mathrm{r}$ or $A_\mathrm{R}$& Rise & Decline (0$-$15\,d, 50$-$100\,d) &  Source(s) \\
     &  &  &   & & days & mag/day ($r$ or $R$)  &  \\
\hline
    SN~2008S & 0.0002& ILRT &  -13.26& 0.740 (R) &7.0 & 0.0185, 0.0378& \citealt{Thompson2009, Adams2016}\\
    NGC~300-2008-OT1 & 0.00048& ILRT& -13.37& 0.027 (R)& 28.3& 0.0118, 0.0298& \citealt{Thompson2009, Humphreys2011, Adams2016}\\ 
    PTF~10fqs & 0.008& ILRT& -11.929& 0.089 (r) &10.80& 0.0535, 0.0332& \citealt{Kasliwal2011}\\ 
    AT~2013la & 0.002712& ILRT& -13.24& 0.024 (r) &7.0& 0.0353, 0.0162& \citealt{Cai2021}\\
    AT~2017be& 0.001438& ILRT& -12.66& 0.123 (r) &10.5& 0.0329, 0.0182& \citealt{Cai2018}\\
    AT~2017jfs& 0.008& LRN& -14.12& 0.055 (r) &2.5& 0.0689, -0.0068& \citealt{Pastorello2019c}\\
    AT~2018aes & 0.00391& ILRT & -12.70& 0.053 (r)&26.7& 0.0181, 0.0660& \citealt{Cai2021}\\
    AT~2018bwo& 0.00156 & LRN& -11.18& 0.048 (r) &13.2& 0.0388, -& \citealt{Blagorodnova2021, Pastorello2022}\\
    AT~2018hso & 0.0039& LRN& -13.06& 0.026 (r) &10.0& 0.0740, -0.0005& \citealt{Cai2019}\\
    AT~2019zhd & -0.001764& LRN& -9.67& 0.142 (r) &61.3& 0.0818, - & \citealt{Pastorello2021a}\\ 
    AT~2021afy & 0.007208& LRN& -12.63& 0.130 (r) &6.0& 0.0167, 0.0415& \citealt{Pastorello2022}\\
    AT~2021biy & 0.002035 & LRN&  -13.08 & 0.039 (r) & 11.53 & 0.0666, -0.0177 & \citealt{Cai2022b}\\
\hline
\end{tabular}
\end{table*}
\onecolumn 
\begin{landscape}
\begin{longtable}{ccccccccccl}
\caption{Optical photometry. $B$-band and $V$-band photometry are given in Vega magnitudes, the rest in AB magnitudes.}
\label{table:opt_phot}\\
\hline
\hline
      UTC Date &     MJD &  Epoch (d) &      $u$ (err) &      $B$ (err) &      $V$ (err) &      $g$ (err) &      $r$ (err) &      $i$ (err) &      $z$ (err) &       Telescope (Instrument) \\
\hline
2022-04-01 & 59670.4 &    2.4 & 19.08 (0.39) &    - &    - &    - &    - &    - &    - &    LCOGT (fa05) \\
2022-04-01 & 59670.5 &    2.5 &    - & 18.44 (0.27) & 17.96 (0.27) &    - &    - &    - &    - &          LCOGT (fa05) \\
2022-04-01 & 59670.6 &    2.6 &    - & 18.42 (0.14) & 17.93 (0.08) & 18.18 (0.05) & 17.74 (0.05) & 17.65 (0.16) &    - &          LCOGT (kb82) \\
2022-04-02 & 59671.0 &    3.0 &  18.7 (0.04) &    - &    - &    - &    - &    - &    - &    LCOGT (fa11) \\
2022-04-02 & 59671.1 &    3.1 &    - &    - &    - &    - & 17.59 (0.04) &    - &    - & NOT (ALFOSC) \\
2022-04-02 & 59671.2 &    3.2 & 18.69 (0.11) &    - &    - &    - &    - &    - &    - &    LCOGT (fa05) \\
2022-04-02 & 59671.4 &    3.4 &    - &    - &    - &  17.9 (0.04) &    - &    - &    - &    ZTF \\
2022-04-02 & 59671.5 &    3.5 &    - &    - &    - & 17.91 (0.02) &    - &    - &    - &          LCOGT (fa05) \\
2022-04-03 & 59672.1 &    4.1 &    - & 18.11 (0.02) & 17.61 (0.03) & 17.88 (0.02) & 17.49 (0.03) & 17.44 (0.02) &    - &          LCOGT (fa05) \\
2022-04-05 & 59674.4 &    6.4 &    - &    - &    - & 17.92 (0.06) &    - &    - &    - &    ZTF \\
2022-04-05 & 59674.5 &    6.5 &    - &    - &    - &    - & 17.61 (0.01) &    - &    - &    ZTF \\
2022-04-06 & 59675.9 &    7.9 &    - &    - &    - &    - & 17.77 (0.04) &    - &    - & NOT (ALFOSC) \\
2022-04-07 & 59676.2 &    8.2 & 18.56 (0.05) & 18.27 (0.03) & 17.87 (0.02) & 18.03 (0.02) & 17.87 (0.03) & 17.74 (0.02) & 17.78 (0.04) &              LT (IO:O) \\
2022-04-07 & 59676.3 &    8.3 &    - &    - &    - & 18.02 (0.08) &    - &    - &    - &    ZTF \\
2022-04-07 & 59676.4 &    8.4 &    - &    - &    - &    - & 17.89 (0.11) &    - &    - &    ZTF \\
2022-04-07 & 59676.9 &    8.9 &    - &    - &  17.98 (0.1) & 18.17 (0.18) & 17.95 (0.07) &  17.83 (0.1) &    - &          Asiago Schmidt (Moravian) \\
2022-04-08 & 59677.0 &    9.0 & 18.71 (0.14) & 18.35 (0.09) &    - &    - &    - &    - &    - &          Asiago Schmidt (Moravian) \\
2022-04-09 & 59678.2 &   10.2 & 18.88 (0.06) & 18.49 (0.03) & 18.08 (0.02) & 18.28 (0.02) & 17.99 (0.02) & 17.96 (0.02) &  18.0 (0.05) &              LT (IO:O) \\
2022-04-09 & 59678.3 &   10.3 &    - &    - &    - & 18.27 (0.05) &    - &    - &    - &    ZTF \\
2022-04-11 & 59680.2 &   12.2 &  19.21 (0.1) &  18.70 (0.06) & 18.25 (0.03) & 18.49 (0.05) & 18.15 (0.03) & 18.18 (0.03) & 18.16 (0.05) &              LT (IO:O) \\
2022-04-12 & 59681.0 &   13.0 &    - &    - &    - &    - &  18.20 (0.09) &    - &    - & NOT (ALFOSC) \\
2022-04-13 & 59682.2 &   14.2 &    - &    - &    - & 18.59 (0.14) &    - &    - &    - &    ZTF \\
2022-04-13 & 59682.3 &   14.3 &    - &    - &    - &    - & 18.29 (0.09) &    - &    - &    ZTF \\
2022-04-15 & 59684.3 &   16.3 &    - &    - &    - &  18.71 (0.1) & 18.32 (0.08) &    - &    - &    ZTF \\
2022-04-18 & 59687.3 &   19.3 &    - &    - &    - &    - &  18.40 (0.04) &    - &    - &    ZTF \\
2022-04-18 & 59687.8 &   19.8 &    - &    - &    - & 18.88 (0.13) &    - &    - &    - &          Asiago Schmidt (Moravian) \\
2022-04-18 & 59687.9 &   19.9 &    - &    - &    - &    - &  18.40 (0.14) & 18.34 (0.13) &    - &          Asiago Schmidt (Moravian) \\
2022-04-20 & 59689.2 &   21.2 &    - & 19.26 (0.05) & 18.58 (0.02) & 18.97 (0.04) & 18.45 (0.02) & 18.33 (0.02) & 18.33 (0.04) &       NOT (ALFOSC) \\
2022-04-20 & 59689.3 &   21.3 &    - &    - &    - &  19.0 (0.11) & 18.48 (0.06) &    - &    - &    ZTF \\
2022-04-21 & 59690.9 &   22.9 & 20.19 (0.17) &  19.40 (0.06) & 18.68 (0.05) & 19.06 (0.08) & 18.52 (0.05) & 18.37 (0.03) & 18.35 (0.04) &              LT (IO:O) \\
2022-04-23 & 59692.9 &   24.9 & 20.43 (0.15) & 19.56 (0.07) & 18.73 (0.05) & 19.09 (0.06) & 18.56 (0.06) &  18.40 (0.03) & 18.36 (0.04) &              LT (IO:O) \\
2022-04-24 & 59693.2 &   25.2 &    - &    - &    - & 19.11 (0.12) &    - &    - &    - &    ZTF \\
2022-04-24 & 59693.3 &   25.3 &    - &    - &    - &    - & 18.58 (0.03) &    - &    - &    ZTF \\
2022-04-26 & 59695.3 &   27.3 &    - &    - &    - & 19.13 (0.12) &    - &    - &    - &    ZTF \\
2022-04-27 & 59696.9 &   28.9 &    - & 19.75 (0.12) & 18.79 (0.07) & 19.17 (0.04) & 18.59 (0.03) & 18.43 (0.03) & 18.32 (0.03) &              LT (IO:O) \\
2022-04-28 & 59697.3 &   29.3 &    - &    - &    - & 19.18 (0.06) &  18.60 (0.07) &    - &    - &    ZTF \\
2022-04-29 & 59698.0 &   30.0 &    - &    - &    - &    - & 18.61 (0.17) &    - &    - & NOT (ALFOSC) \\
2022-04-30 & 59699.2 &   31.2 &    - &    - &    - & 19.18 (0.05) &    - &    - &    - &    ZTF \\
2022-05-02 & 59701.1 &   33.1 &    - &    - &    - &    - & 18.67 (0.05) &    - &    - & NOT (ALFOSC) \\
2022-05-02 & 59701.3 &   33.3 &    - &    - &    - & 19.18 (0.11) &    - &    - &    - &    ZTF \\
2022-05-02 & 59701.9 &   33.9 & 20.84 (0.18) & 19.87 (0.05) & 18.88 (0.04) & 19.31 (0.04) & 18.67 (0.04) & 18.44 (0.05) & 18.34 (0.03) &              LT (IO:O) \\
2022-05-06 & 59705.9 &   37.9 &    - & 19.99 (0.06) &  19.0 (0.05) & 19.45 (0.05) & 18.73 (0.04) & 18.48 (0.02) & 18.35 (0.03) &              LT (IO:O) \\
2022-05-10 & 59709.2 &   41.2 &    - &    - &    - &    - & 18.71 (0.09) &    - &    - &               ZTF \\
2022-05-10 & 59709.3 &   41.3 &    - &    - &    - &    - & 18.77 (0.12) &    - &    - &               ZTF \\
 \hline
 \newpage
 \caption{Continued.}
 \\
 \hline
 \hline
       UTC Date &     MJD &  Epoch (d) &      u (err) &      B (err) &      V (err) &      g (err) &      r (err) &      i (err) &      z (err) &       Telescope (Instrument) \\
\hline
2022-05-12 & 59711.3 &   43.3 &    - &    - &    - &    - & 18.74 (0.13) &    - &    - &               ZTF \\
2022-05-12 & 59711.9 &   43.9 &    - &  20.18 (0.2) &  19.20 (0.06) & 19.66 (0.08) &  18.80 (0.04) & 18.59 (0.03) & 18.37 (0.04) &              LT (IO:O) \\
2022-05-16 & 59715.2 &   47.2 &    - &    - &    - & 19.71 (0.13) &    - &    - &    - &               ZTF \\
2022-05-19 & 59718.2 &   50.2 &    - &    - &    - &    - &  18.97 (0.1) &    - &    - &               ZTF \\
2022-05-20 & 59719.9 &   51.9 &    - & 20.45 (0.09) & 19.38 (0.05) & 19.93 (0.05) & 19.04 (0.04) & 18.74 (0.05) & 18.54 (0.04) &              LT (IO:O) \\
2022-05-24 & 59723.9 &   55.9 &    - & 20.69 (0.17) & 19.61 (0.13) & 20.15 (0.09) & 19.25 (0.04) & 18.91 (0.05) &  18.74 (0.1) &              LT (IO:O) \\
2022-05-25 & 59724.9 &   56.9 &    - &    - &    - &    - & 19.29 (0.04) &    - &    - &            GTC (OSIRIS) \\
2022-06-02 & 59732.0 &   64.0 &    - & 21.04 (0.09) & 19.95 (0.04) & 20.41 (0.05) & 19.53 (0.03) & 19.18 (0.03) & 18.92 (0.06) &              LT (IO:O) \\
2022-06-09 & 59739.9 &   71.9 &    - & 21.67 (0.37) & 20.44 (0.16) & 20.98 (0.12) & 19.93 (0.08) & 19.52 (0.04) & 19.15 (0.07) &              LT (IO:O) \\
2022-06-17 & 59747.0 &   79.0 &    - & 22.34 (0.18) & 20.83 (0.06) & 21.45 (0.14) &  20.30 (0.04) & 19.89 (0.03) & 19.48 (0.04) &       NOT (ALFOSC) \\
2022-06-18 & 59748.0 &   80.0 & 23.42 (0.47) &    - &    - &    - &    - &    - &    - &       NOT (ALFOSC) \\
2022-06-22 & 59752.0 &   84.0 &    - & 22.68 (0.22) & 21.19 (0.11) &  21.79 (0.1) & 20.59 (0.09) & 20.05 (0.05) & 19.72 (0.07) &       NOT (ALFOSC) \\
2022-06-26 & 59756.9 &   88.9 &    - &    - & 21.56 (0.15) & 22.24 (0.26) & 20.87 (0.06) & 20.36 (0.05) &  20.10 (0.07) &       NOT (ALFOSC) \\
2022-07-11 & 59771.9 &  103.9 &    - &    - &    - &    - &  21.70 (0.39) &    - &    - &              LT (IO:O) \\
2022-07-12 & 59772.0 &  104.0 &    - &    - &    - &    - &    - &  21.20 (0.22) & 20.75 (0.23) &              LT (IO:O) \\
2022-07-13 & 59773.9 &  105.9 &    - &    - & 22.55 (0.24) &  23.42 (0.4) & 21.78 (0.17) &    - &    - &       NOT (ALFOSC) \\
2022-07-18 & 59778.9 &  110.9 &    - &    - &    - &    - & 22.26 (0.18) & 21.41 (0.18) & 20.81 (0.17) &       NOT (ALFOSC) \\
2022-07-31 & 59791.9 &  123.9 &    - &    - &    - &    - &    - & 21.84 (0.44) & 21.04 (0.15) &       NOT (ALFOSC) \\
2022-08-06 & 59797.9 &  129.9 &    - &    - &    - &    - &    - &  22.20 (0.26) & 21.26 (0.17) &       NOT (ALFOSC) \\
2022-08-15 & 59806.9 &  138.9 &    - &    - &    - &    - &    - & 22.84 (0.46) &    - &       NOT (ALFOSC) \\
\hline
\end{longtable}
\end{landscape}
\twocolumn 

\begin{table*}
\caption{NIR photometry given in Vega magnitudes.}
\centering
\begin{tabular}{ccccccl}
\hline
\hline
      UTC Date &     MJD &  Epoch &      $J$ (err) &      $H$ (err) &     $K_{\mathrm{s}}$ (err) & Telescope (Instrument) \\
\hline
2022-05-07 & 59706.0 &  38.0 & 17.46 (0.04) &    17.11 (0.06) &    16.87 (0.04) &     NOT (NOTCam) \\
2022-06-01 & 59731.0 &  63.0 & 17.84 (0.04) & 17.35 (0.06) & 17.02 (0.06) &     NOT (NOTCam) \\
2022-07-09 & 59769.9 & 101.9 & 18.94 (0.09) & 17.84 (0.06) & 17.42 (0.09) &     NOT (NOTCam) \\
2022-07-26 & 59786.9 & 118.9 &  19.16 (0.10) & 18.08 (0.09) & 17.59 (0.09) &     NOT (NOTCam) \\
2022-08-17 & 59808.9 & 140.9 &    - &    - & 17.63 (0.12) &     NOT (NOTCam) \\
\hline
\end{tabular}
\label{table:NIR_phot}
\end{table*}

\begin{table*}
\caption{Optical spectroscopy. Instrumental resolutions were determined based on measurement of the [O~{\sc i}] $\lambda$5577.34 skyline except in the case of those marked with an asterisk, where reference values were used.} 
\centering
\label{table:opt_spec}
\begin{tabular}{ccccccl}
\hline
\hline
      UTC Date &     MJD &  Epoch (d) &  Exp. (s) & Slit ($''$) & $\Delta\lambda$ (\AA) &Telescope (Instrument, Grism) \\
\hline
2022-04-01 & 59670.4
 &   2.4      &    3000     &   2.1       & 7.83* &           APO 3.5m (KOSMOS) \\ 
2022-04-02 & 59671.2 &        3.2 &           3000 &            1.3 &  17.49 &           NOT (ALFOSC, \#4) \\
2022-04-06 & 59675.9 &        7.9 &           3000 &            1.0 &  12.87  &          NOT (ALFOSC, \#4) \\
2022-04-11 & 59681.0 &       13.0 &           2467 &            1.0 & 12.84   &          NOT (ALFOSC, \#4) \\
2022-04-22 & 59691.1 &   23.1      &   3600    &    1.0     &   9.37*   &        
              TNG (DOLORES, LR-B + LR-R)\\ 
2022-05-02 & 59701.1 &       33.1 &           3600 &            1.3 & 17.17  &           NOT (ALFOSC, \#4) \\
2022-05-25 & 59724.9  &  56.9       &   3600       &   1.0   &  9.75*  &  GTC (OSIRIS,  R1000R) \\
2022-05-27 & 59726.1 &   58.9      &   3600     &   1.5    & 13.79*  &              
              TNG (DOLORES, LR-R) \\  %
2022-06-28  &59758.3  &  90.3       &  3600       &    1.0  &  5.43*  &             Keck (LRIS, 600/4000 + 400/8500) \\
\hline
\end{tabular}
\end{table*}
\end{appendix}
\end{document}